# Full-Duplex eNodeB and UE Design for 5G Networks

*Abstract*—The recent progress in the area of self-interference cancellation (SIC) design has enabled the development of full-duplex (FD) single and multiple antenna systems. In this paper, we propose a design for FD eNodeB (eNB) and user equipment (UE) for 5G networks. The use of FD operation enables simultaneous in-band uplink and downlink operation and thereby cutting down the spectrum requirement by half. FD operation requires the same subcarrier allocation to UE in both uplink and downlink. Long Term Evolution (LTE) uses orthogonal frequency division multiple access (OFDMA) for downlink. To enable FD operation, we propose using single-carrier frequency division multiple access (SC-FDMA) for downlink along with the conventional method of using it for uplink. Taking advantage of channel reciprocity, singular value decomposition (SVD) based beamforming in the downlink allows multiple users (MU) to operate on same set of subcarriers. In uplink, frequency domain minimum mean square error (MMSE) equalizer along with successive interference cancellation with optimal ordering (SSIC-OO) algorithm is used to decouple signals of users operating in the same set of subcarriers. The work includes simulations showing efficient FD operation both at UE and eNB for downlink and uplink respectively.

## I. INTRODUCTION

The Wireless World Research Forum (WWRF) predicted that 7 trillion wireless devices will serve 7 billion people worldwide by 2017 [1]. The explosion in the number of mobile users has resulted in the depletion of limited spectrum resource. With 4G getting deployed or soon to be deployed in many countries, this problem of spectrum crisis can prove to be a bottleneck. To address this issue, research has already began on 5G which is expected to get deployed beyond 2020 [1]. One of the major revolutions can be the introduction of full-duplex (FD) eNodeB (eNB) and user equipment (UE). The FD operation has the capability of cutting down the spectrum requirement by half.

In traditional FDD, two separate channels are used for uplink and downlink. FD systems make the in-band transceiving feasible. This means using a single channel simultaneously for uplink and downlink without sacrificing any temporal resource for both the links, which is the case for TDD operation. In recent years, excessive work is being done in the area of self-interference cancellation (SIC) design for both single and multiple antenna transceiver units [2-4]. This enables optimal cancellation of interference from the receiver chain introduced by the transmitter chain (or chains) of the transceiver unit. In this paper, we discuss the changes in uplink and downlink operation required for incorporating FD into cellular networks and propose the corresponding revamp for eNB and UE.

In conventional LTE system, users are allocated sub carrier resources according to channel state scheduling algorithm [5]. For uplink and downlink, single carrier frequency division multiple access (SC-FDMA) and orthogonal frequency division multiple access (OFDMA) is used for multiple access respectively [6]. For FD operation, same spectrum resource (, i.e., subcarriers) needs to be allocated to users for uplink and downlink. Hence we propose using SC-FDMA for both uplink and downlink. Besides achieving better bit error rate (BER) than OFDMA at the UE, SC-FDMA allows design of energy efficient 'green' eNB because of its low peak-to-average power ratio (PAPR) [7].

Another revolutionary technology for 5G networks is the Massive MIMO technology which supports hundreds of antennas at the eNB [8, 9]. The FD operation enables channel reciprocity for uplink and downlink. This eases the availability of channel state information (CSI) at the transmitter. This coupled with the Massive MIMO technology allows multiple user (MU) to operate on the same set of subcarriers. Frequency domain MMSE equalizer along with successive interference cancellation with optimal ordering (SSIC-OO) and singular value decomposition (SVD) based beamforming allow MU subcarrier sharing both for uplink and downlink respectively.

The paper is divided into six sections. In section 2, we have discussed about the system model describing the eNB and UE design for the proposed method. The FD architecture used for designing the modified eNB and UE is briefly analyzed in section 3. Section 4 deals with uplink and downlink operation for the FD cellular network. Simulation results for downlink and uplink are discussed in section 5. The conclusion is presented in Section 6.

*Notation:* $[.]^T, (.)^H$ denote transpose and Hermitian respectively. $\|.\|_2$ denotes euclidean norm.

## II. SYSTEM MODEL

The SIC cancellation design deployed at the RF front end of both eNB and UEs will allow single channel for both uplink and downlink operation. The SIC design requires additional digital and analog taps which increase the hardware complexity drastically. This complexity increases linearly, with the number of antennas, in SIC design of [2,4]. The complexity is a major limitation when comes to deployment in multiple antenna UE where rapid depletion of battery power is also a major issue. The SIC design is analyzed in the next section. In this work, we consider an eNB with $N_e$ antennas and $K$ single antenna UE (category one) operating in FD mode as shown in Fig.1.

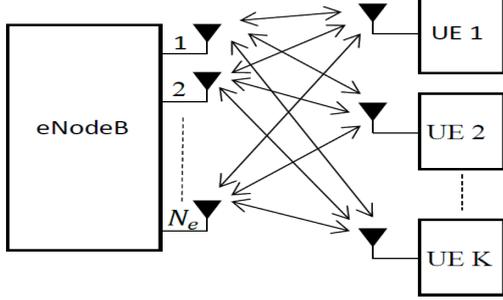

Fig. 1: Multiuser full-duplex system model

We assume dynamic subcarrier allocation based on channel state information. Let each subcarrier allocated be shared by $K'$ users simultaneously, where $K'$ is given as $min(N_e,K)$. Hence each user is allocated $M$ $(=\lfloor \frac{NK'}{K} \rfloor)$ subcarriers. The recent development in the area of Massive MIMO technology allows deployment of hundreds of antennas at the eNB. One of the commercially available Massive MIMO design to support 5G is TitanMIMO by Nutaq [8]. Keeping this in mind we here consider a case of $K'=K$, i.e., all the $K$ users are allocated all the $N$ subcarriers. In actual practice, the users are allocated $N'$ $(< N)$ subcarriers as not all of the $N$ subcarriers are used as data subcarriers. The channel between each eNB antenna and UEs antenna is assumed to be frequency selective with $L$ taps. The FD allows the channel reciprocity between downlink and uplink:

$$h_{j,i}^u(b) = h_{j,i}^d(b) \qquad (1)$$

where $b = 0, 1, 2, ..., L-1$, $h_{j,i}^u(b)$ and $h_{j,i}^d(b)$ denotes $b^{th}$ time domain uplink and downlink channel coefficient between $j^{th}$ antenna at eNB and antenna of $i^{th}$ UE respectively. The proposed FD eNB and UE design is shown in the Fig.2 and Fig.3 respectively.

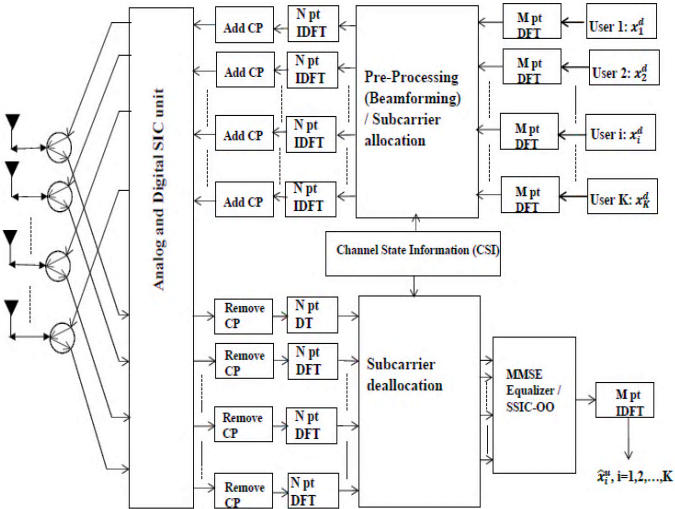

Fig. 2: Transceiver structure for the proposed eNB architecture

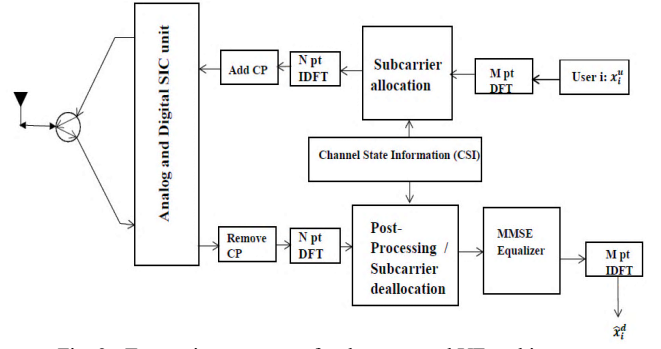

Fig. 3: Transceiver structure for the proposed UE architecture

Let us now look at the primary features in transmit and receive chain for each uplink and downlink operation which enable MU spatial multiplexing. In the uplink, frequency domain MMSE equalization is done along successive interference cancellation with optimal ordering (SSIC-OO) algorithm at the eNB. In downlink, the channel reciprocity property of FD enables the transmitter (, .i.e., eNB) to acquire CSI with ease. The CSI can be used to perform efficient subcarrier allocation and precoding the user data so as to perform SVD based beamforming. It is seen that all the complex signal processing is done at the eNB. This frees the UEs of additional computational complexity. Each of these is thoroughly discussed in the section 4. In this work, to keep the analysis for the FD system simple, we have used frequency domain MMSE equalizer for equalization due to its low complexity. However, more accurate and complex equalization and detection methods [7,10], can be used to get improved performance.

### III. ANALYSIS OF FULL DUPLEX ARCHITECTURE

#### A. Current research on Full-Duplex (FD) systems

Recent work [2-4] in FD systems have invalidated the long-held assumption that the communication can only be done in the half-duplex mode. Application of the FD system allows high bandwidth applications through carrier aggregation and solves the spectrum fragmentation problem which can create significant hindrance in 5G network evolution [3]. *For facilitating FD communication, both eNB and UE must operate in FD mode.* In practical scenarios, the transmitted signal is a nonlinear function of the ideal transmitted signal [2]. This transformation results when the digital baseband signal is converted to analog and up-converted to the carrier frequency. Various analog components involved in this operation, e.g. power amplifier and analog circuits, introduce linear and nonlinear distortion and also add noise (transmit noise and phase noise). These factors prevent direct subtraction of transmitted signal from the received signal for the purpose of self-interference cancellation (SIC) for in-band FD operation [2].

To compensate above mentioned distortions at the receiver, any SIC architecture aims to model and cancel them out. The recent noteworthy contribution toward the practical implementation of the FD system can be found in [2,4]. In [2], demonstrates implementation of FD operation for a single antenna system and complete cancellation of self-interference to the noise floor. The work is extended to multiple antenna systems in [4]. In our work, the proposed architecture involves the SIC cancellation design for single and multiple antenna system incorporated in UEs and eNB design respectively. In the subsequent sub-section, we briefly review the implications of SIC design in our proposed architecture. The practical implementation of the FD systems in eNB transmitting with high power and mobile UE is challenging and needed to be examined more thoroughly, however, details of the SIC design are not discussed here.

### B. Use of FD technology in proposed architecture

The self-interference at a receive chain of multiple antenna eNB consists of 1)"self-talk", .i.e., interference from the transmit chain with which receive chain shares antenna and 2) "cross-talk", .i.e., interference from neighbouring transmit chains. There are three major components for self-talk and cross talk [4]: 1) Linear component, 2) Non-linear component and 3) Transmit Noise. The SIC design dynamically model the self-talk and cross-talk (in case of multiple antenna system, here eNB) components introduced at the baseband signal in the transmit chain and cancel them from the receive chain. In single antenna UE, self-interference just refers to self-talk.

There are two stages for SIC cancellation. The first stage is analog cancellation which implements SIC in the analog domain and prevents saturation of receive chain. The analog cancellation however does not cancel all the self-interference. Hence, in the second stage, it is followed by digital cancellation which cancels out the remaining self-interference. For the proposed architecture shown in Fig.2 and Fig.3, the *Analog and Digital SIC unit*, includes both analog and digital cancellation stage [2,4]. The complexity of SIC design described in [4] scales linearly with number of antennas used. However, the residual interference remains equal to a single antenna system.

### IV. UPLINK AND DOWNLINK OPERATION

#### A. Analysis of Uplink Operation

Let $x_i^u$ denotes the information data block of length $M$ for $i^{th}$ user in uplink (denoted by $u$):

$$x_i^u = [x_i^u(1), x_i^u(2),..., x_i^u(M)]^T, i = 1,2,...,K \quad (2)$$

The output of the $M$-point DFT block (Fig.3.) is given by:

$$\overline{x}_i^u = F_M x_i^u \quad (3)$$

where $F_M$ is the $M$-point DFT matrix, $\overline{x}_i^u = [\overline{x}_i^u(1), \overline{x}_i^u(2),..., \overline{x}_i^u(M)]^T$. Let $A^i$, represents the $N \times M$ subcarrier allocation matrix for $i^{th}$ user, $i=1,2,...,K$ [7]. *Due to channel reciprocity, the subcarrier allocation matrix for a user in both uplink and downlink is equal.* The $N \times 1$ vector input to the $N$-point IDFT block for $i^{th}$ user is given by:

$$d_i^u = A^i \overline{x}_i^u \quad (4)$$

The output of $N$-point IDFT block for the $i^{th}$ user is given by:

$$s_i^u = \overline{F}_N d_i^u \quad (5)$$

where $\overline{F}_N$ denotes the N-point IDFT matrix. This signal is then transmitted from the antenna of the $i^{th}$ UE after addition of cyclic prefix (CP).

At the eNB, the received signal is obtained after SIC cancellation in the $N_e$ receive chains. The received signal vector of size $N \times 1$ at the $j^{th}$ receive antenna due to the $K$ users, after removing the CP, is given by:

$$y_j^u = \sum_{i=1}^{K} h_{j,i}^u \otimes s_i^u + n_j, j = 1,2,...,N_e \quad (6)$$

where $\otimes$ denotes $N$-point circular convolution operation, $h_{j,i}^u = [h_{j,i}^u(0), h_{j,i}^u(1),..., h_{j,i}^u(L-1), (N-L)zeros]^T$, and $n_j \in cN(0, N_0 I_N)$ is the additive noise vector, which due to channel reciprocity, is equal for each pair of antenna in eNB and UE in both uplink and downlink. The output of the $j^{th}$ antenna received signal is then converted to the frequency domain by taking $N$-point DFT, which is given by:

$$\overline{y}_j^u = F_N y_j^u, j = 1,2,...,N_e \quad (7.1)$$

$$\overline{y}_j^u = \sum_{i=1}^{K} H_{j,i}^u d_i^u + \overline{n}_j \quad (7.2)$$

where $H_{j,i}^u = diag(F_N h_{j,i}^u)$ is the $N \times N$ diagonal matrix whose diagonal elements are the frequency domain channel coefficients between antenna of $i^{th}$ UE and $j^{th}$ receive antenna at eNB. Let $\overline{A}^i$ be the $M \times N$ deallocation matrix where $\overline{A}^i = (A^i)^T$. The $M$-point received signal on the $j^{th}$ antenna after sub-carrier deallocation is given by:

$$\widetilde{y}_j^u = \overline{A}^i \overline{y}_j^u \quad (8)$$

For the $m^{th}$ subcarrier, .i.e., $m^{th}$ element of $\widetilde{y}_j^u$, the received signal on the $j^{th}$ antenna is given by:

$$\widetilde{y}_j^u(m) = \sum_{i=1}^{K} H_{j,i}^u(m) d_i^u(m) + \widetilde{n}_j(m) \quad (9)$$

The signal received by the eNB on all the $N_e$ antennas for the $m^{th}$ subcarrier is given by:

$$\widetilde{y}^u(m) = \sum_{i=1}^{K} H_i^u(m) d_i^u(m) + \widetilde{n}(m), i = 1,2,...,K \quad (10)$$

where $\widetilde{y}^u(m) = [\widetilde{y}_1^u(m), \widetilde{y}_2^u(m),..., \widetilde{y}_{N_e}^u(m)]^T$,
$H_i^u(m) = [H_{1,i}^u(m), H_{2,i}^u(m),..., H_{N_e,i}^u(m)]^T$, and
$\widetilde{n}(m) = [\widetilde{n}_1(m), \widetilde{n}_2(m),..., \widetilde{n}_{N_e}]^T$

For decoding of $l^{th}$ user signal, the signal received given by *(10)* can be represented as:

$$\widetilde{y}^u(m) = H_l^u(m) d_l^u(m) + \sum_{\substack{i=1 \\ i \neq l}}^{K} H_i^u(m) d_i^u(m) + \widetilde{n}(m) \quad (11)$$

where the first term represents the desired user signal, the second term represents the co-channel interference from the other users and the last term is the noise term.

The received signal is subsequently passed through the frequency domain MMSE equalizer. The estimated signal for the $l^{th}$ user on the $m^{th}$ subcarrier is given by:

$$\hat{d}_l^u(m) = [\sigma_{d_l^u(m)}^{-1} + (H_l^u(m))^H R_{\widetilde{q}\widetilde{q}}^{u^{-1}}(m) H_l^u(m)]^{-1} (H_l^u(m))^H R_{\widetilde{q}\widetilde{q}}^{u^{-1}}(m) \widetilde{y}(m) \quad (12)$$

where $\tilde{q}^u(m) = \sum_{\substack{i=1 \\ i \neq l}}^{K} H_l^u(m)d_i^u(m) + \tilde{n}(m)$ is the co-channel interference and noise factor for the $l^{th}$ user. $R_{\tilde{q}\tilde{q}}^u(m) = \sum_{\substack{i=1 \\ i \neq l}}^{K} H_l^u(m)(H_l^u(m))^H \sigma_{d_l^u(m)}^2 + N_0 I_{N_e}$ is the covariance of $\tilde{q}^u(m)$. $\sigma_{d_l^u(m)}^2$ is the signal power for the $l^{th}$ user. The signal power is normalized to have unit power, i.e., $\sigma_{d_l^u(m)}^2 = 1$, hence the estimated signal term for the $l^{th}$ user on the $m^{th}$ subcarrier is given by:

$$\hat{d}_l^u(m) = [1 + (H_l^u(m))^H R_{\tilde{q}\tilde{q}}^{u^{-1}}(m) H_l^u(m)]^{-1} (H_l^u(m))^H R_{\tilde{q}\tilde{q}}^{u^{-1}}(m)\tilde{y}(m) \quad (13)$$

Now ignoring the scaling term, (14) can be given by:

$$\hat{d}_l^u(m) = (H_l^u(m))^H R_{\tilde{q}\tilde{q}}^{u^{-1}}(m)\tilde{y}(m) \quad (14)$$

The estimated signal of the $l^{th}$ user can be used to cancel out its effect from the overall received signal. The signal then can be used to estimate the signal of other users. This process is repeated till received signals from the all the users are estimated. The order in which the signal is estimated is determined by the receive signal power of the users. We below define the algorithm called successive interference cancellation with optimal ordering (SSIC-OO) for this purpose:

**Algorithm: SSIC-OO for estimating user signal**
1. **Let** $c=K$
2. **while**($c>1$) **do**
3. Calculate received power for all the $K$ users $PW_i = \|H_i^u(m)\|^2, i=1,2,...,K$ where $\sigma_{d_l^u(m)}^2 = 1$
4. Let $l = \arg\max_i(PW_i)$
5. Estimate $\hat{d}_l^u(m)$
6. $\tilde{y}^u(m) = \tilde{y}^u(m) - H_l^u(m)\hat{d}_l^u(m)$
7. $K=K-1$
8. $c=c-1$
9. **end while**
10. Now let $l$ represent index for the last remaining unestimated user
11. For MMSE equalizer, ignore $R_{\tilde{q}\tilde{q}}^{u^{-1}}(m)$ which is now just a scaling factor: $R_{\tilde{q}\tilde{q}}^{u^{-1}}(m) = [N_o I_{N_e}]^{-1}$
12. $\hat{d}_l^u(m) = (H_l^u(m))^H \tilde{y}_l^u(m)$, the system represents a single user and multiple receive antennas at eNB with maximal ratio combining (MRC) of user symbol

Let the signal for the $l^{th}$ user on all the subcarrier is given by:

$$\hat{d}_l^u = [\hat{d}_l^u(1), \hat{d}_l^u(2),..., \hat{d}_l^u(M)]^T \quad (15)$$

This signal for the $l^{th}$ user is then converted to time domain by an $M$-point IDFT operation given by:

$$\hat{x}_l^u = \overline{F}_M \hat{d}_l^u, l=1,2,...,K \quad (16)$$

where $\overline{F}_M$ is $M$-point inverse IDFT matrix. This signal is then used for decoding of the signal for the $l^{th}$ user. The decoding procedure is out of scope of this paper.

*B. Analysis of FD Downlink Operation*

Let $x_i^d$ denote the $i^{th}$ user information data block of length $M$ in downlink (denoted by $d$):

$$x_i^d = [x_i^d(1), x_i^d(2),..., x_i^d(M)]^T, i=1,2,...,K \quad (17)$$

The output of the $M$-point block (Fig.3) is given by:

$$\overline{x}_i^d = F_M x_i^d \quad (18)$$

where $F_M$ is the $M$-point FDT matrix and $\overline{x}_i^d = [\overline{x}_i^d(1), \overline{x}_i^d(2),..., \overline{x}_i^d(M)]^T$. This output is then passed through the beamforming and subcarrier allocation block. Let $P_m^i$ denotes $N_e \times 1$ beamforming/precoding vector for $i^{th}$ user on $m$th subcarrier. The precoded output vector of size $N_e \times 1$ for $i^{th}$ user on $m^{th}$ subcarrier is given by:

$$z_i^d(m) = P_m^i \overline{x}_i^d(m), m=1,2,...,M \quad (19)$$

where $z_i^d(m) = [z_{i,1}^d(m), z_{i,2}^d(m),..., z_{i,N_e}^d(m)]^T$. As discussed, $A^i$ represents the $N \times M$ subcarrier allocation matrix for $i^{th}$ user, $i=1,2,...,K$. The $N \times 1$ vector input to the $N$-point IDFT block for $j^{th}$ transmit chain is given by:

$$e_j^d = \sum_{i=1}^{K} A^i z_{i,j}^d, j=1,2,...,N_e \quad (20)$$

where $z_{i,j}^d = [z_{i,j}^d(1), z_{i,j}^d(2),..., z_{i,j}^d(M)]^T$. The output of the $N$-point IDFT block for the $j^{th}$ transmit chain is given by:

$$s_j^d = \overline{F}_N e_j^d, j=1,2,..., N_e \quad (21)$$

where $\overline{F}_N$ denotes the $N$-point IDFT matrix. This signal is then transmitted on $j^{th}$ antenna after addition of the CP.

At the $l^{th}$ UE, the received signal is obtained after SIC cancellation in receive chain. The received signal, after removing the CP, is given by:

$$y_l^d = \sum_{j=1}^{N_e} h_{j,l}^d \otimes s_j^d + \sum_{\substack{m=1 \\ m \neq l}}^{K} h_{m,l}^u \otimes s_m^u + n_l, l=1,2,...,K \quad (22)$$

where the first term is the signal received from the eNB, second term is the interference on the $l^{th}$ UE from the uplink signals of rest of $(K-1)$ UEs and the third term is the channel noise. The second term, where $h_{m,l}^u$ denotes the channel coefficient between the $(K-1)^{th}$ UEs and the $l^{th}$ UE and $s_m^u$ are the uplink signal transmitted by $(K-1)^{th}$ UEs, results from the fact that the $K$ UEs are sharing the same spectrum resource in uplink and downlink. However, in a practical scenario, a plethora of man-made and natural obstructions are present, like buildings and trees, between the UEs since UEs are at a lower altitude. This leads to screening of signals between the UEs (Fig.4). This issue is absent for connection between eNB and UEs by placing eNB at a considerably higher altitude. *Hence by assuming the UEs are sufficiently separated from each user and taking in account the effects like reflection, diffraction, absorption and shadowing* resulting in screening effect and multipath fading, the interference from the (K-1) UEs on $l^{th}$ UE is much reduced in strength. Hence the interference term has been assumed to have minimum effect on the received signal at the $l^{th}$ UE and hence neglected.

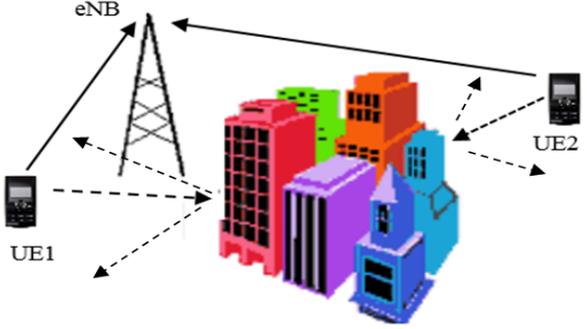

Fig. 4: Screening effect between UEs in urban scenario during uplink

This assumption can be obviated by the use of multiple antennas at UEs. Though it increases complexities at UEs as discussed earlier, it results in improving QoS. By applying appropriate scaling at each antenna, strong beam is steered in the direction of eNB and form nulls in the direction of other interfering UEs [11]. *This multiple antenna UE methodology is beyond the scope of this paper and carried out as next part of the present work.*

Hence, considering the single antenna UE scenario, the received signal vector of size $N \times 1$ from (22) for the $l^{th}$ UE is given by:

$$y_l^d = \sum_{j=1}^{N_e} h_{j,l}^d \otimes s_j^d + n_l, l=1,2,...,K \quad (23)$$

where $\otimes$ denotes N-point circular convolution operation, $h_{j,l}^d = [h_{j,l}^d(0), h_{j,l}^d(1),..., h_{j,l}^d(L-1), (N-L)zeros]^T$, and $n_l \in cN(0, N_0 I_N)$ is the additive noise vector. This signal is then converted to the frequency domain by taking N-point DFT, which is given by:

$$\overline{y}_l^d = F_N y_l^d \quad (24)$$

where $F_N$ is the N-point DFT matrix.

$$\overline{y}_l^d = \sum_{j=1}^{N_e} H_{j,l}^d e_j^d + \overline{n}_l \quad (25.1)$$

$$= \sum_{j=1}^{N_e} H_{j,l}^d \sum_{i=1}^{K} A_i^d z_{i,j}^d + \overline{n}_l \quad (25.2)$$

where $H_{j,l}^d = diag(F_N h_{j,l}^d)$ is the $N \times N$ diagonal matrix whose diagonal elements are frequency domain coefficients between $j^{th}$ transmit antenna at eNB and receive antenna of $l^{th}$ UE. As discussed, $\overline{A}^i$ is the $M \times N$ deallocation matrix where $\overline{A}^i = (A^i)^T$. The M-point received signal for $l^{th}$ user after sub-carrier deallocation is given by:

$$\widetilde{y}_l^d = \overline{A}^i \overline{y}_l^d \quad (26.1)$$

$$= \sum_{j=1}^{N_e} \sum_{i=1}^{K} \overline{A}^i H_{j,l}^d A^i z_{i,j}^d + \widetilde{n}_l \quad (26.2)$$

Now, the received signal for the $l^{th}$ user on the $m^{th}$ subcarrier is given by:

$$\widetilde{y}_l^d(m) = H_l^d(m) \sum_{i=1}^{K} P_m^i x_i^d(m) + \widetilde{n}_l(m) \quad (27)$$

where $H_l^d(m)$ is the $1 \times N_e$ frequency domain channel coefficient vector of $l^{th}$ user on the $m^{th}$ subcarrier. $(1,j)^{th}$ entry is the $m^{th}$ diagonal element of matrix $H_{j,l}^d$. The SVD decomposition of channel matrix $H_l^d(m)$ is given by[12]:

$$H_l^d(m) = U_{m,l}^d \Sigma_{m,l}^d (V_{m,l}^d)^H \quad (28)$$

where for a single antenna UE, $U_{m,l}^d$ is a scalar such that $(U_{m,l}^d)^2 = 1$, $\Sigma_{m,l}^d$ is a scalar equal to $(\lambda_{m,l}^d)^{1/2}$ where $\lambda_{m,l}^d$ is the eigenvalue of $H_l^d(m)(H_l^d(m))^H$ and $V_{m,l}^d$ is a $N_e \times 1$ matrix containing the eigenvector corresponding to non-zero eigenvalue of $(H_l^d(m))^H H_l^d(m)$, which is equal to $\lambda_{m,l}^d$. Let us define the following vectors and matrices:

$$\widetilde{x}^d(m) \doteq [\widetilde{x}_1^d(m), \widetilde{x}_2^d(m),..., \widetilde{x}_K^d(m)]^T$$
$$P_m \doteq [P_m^1, P_m^2,..., P_m^K]$$
$$U_m^d \doteq diag(U_{m,1}^d, U_{m,2}^d,..., U_{m,K}^d)$$
$$\Sigma_m^d \doteq diag(\Sigma_{m,1}^d, \Sigma_{m,2}^d,..., \Sigma_{m,K}^d)$$
$$V_m^d \doteq [V_{m,1}^d, V_{m,2}^d,..., V_{m,K}^d]$$

The received signal vector on $m^{th}$ due to all users sharing the subcarriers is hence can be given by:

$$\widetilde{y}^d(m) = U_m^d \Sigma_m^d (V_m^d)^H P_m \widetilde{x}^d(m) + \widetilde{n}(m) \quad (29)$$

where $\widetilde{y}^d(m) = [\widetilde{y}_1^d(m), \widetilde{y}_2^d(m),..., \widetilde{y}_K^d(m)]^T$, $\widetilde{n}(m) = [\widetilde{n}_1(m), \widetilde{n}_2(m),..., \widetilde{n}_K(m)]^T$

The interference from other users on the $l^{th}$ user can be completely eliminated by choosing the beamforming/precoding matrix as:

$$P_m = [(V_m^d)^H]^+ \alpha_m \quad (30)$$

Where $[(V_m^d)^H]^+$ pseudo inverse of is $V_m^d$, $\alpha_m = diag(\alpha_m^1, \alpha_m^2,..., \alpha_m^K)$ defines the optimal power allocated to each of the K users on each subcarrier. It will be discussed later. The equation (29) can be represented as:

$$\widetilde{y}^d(m) = U_m^d \Sigma_m^d \alpha_m \widetilde{x}^d(m) + \widetilde{n}(m) \quad (31)$$

The received signal on $m^{th}$ subcarrier for the $l^{th}$ user is given by:

$$\widetilde{y}_l^d(m) = U_{m,l}^d \widetilde{\Sigma}_{m,l}^d \widetilde{x}_l^d(m) + \widetilde{n}_l(m) \quad (32)$$

where $\widetilde{\Sigma}_{m,l}^d = \Sigma_{m,l}^d \alpha_m^l$

In the post-processing unit, for the $l^{th}$ user, the received signal is multiplied with $(U_{m,l}^d)^H$:

$$\hat{y}_l^d(m) = (U_{m,l}^d)^H \widetilde{y}_l^d(m) \quad (33.1)$$
$$= \widetilde{\Sigma}_{m,l}^d \widetilde{x}_l^d(m) + w_l(m) \quad (33.2)$$

where for a single antenna UE, $(U_{m,l}^d)^H = U_{m,l}^d$

Let define the following vectors and matrices:
$$\hat{y}_l^d \doteq [\hat{y}_l^d(1), \hat{y}_l^d(2),..., \hat{y}_l^d(M)]^T$$
$$\widetilde{x}_l^d \doteq [\widetilde{x}_l^d(1), \widetilde{x}_l^d(2),..., \widetilde{x}_l^d(M)]^T$$

$$\widetilde{\Sigma}_l^d \doteq diag\ (\Sigma_{1,l}^d, \widetilde{\Sigma}_{2,l}^d,..., \widetilde{\Sigma}_{M,l}^d)$$
$$w_l \doteq [w_l(1), w_l(2),..., w_l(M)]^T$$

Using the above definition, the received signal vector for $l^{th}$ user on the $M$ allocated subcarriers is given by:

$$\hat{y}_l^d = \widetilde{\Sigma}_l^d\ \widetilde{x}_l^d + w_l \quad (34)$$

This is then subjected to frequency domain MMSE equalization. The received signal vector at the output the MMSE equalizer on the $M$ allocated subcarriers is:

$$\hat{\widetilde{x}}_l^d = ((\widetilde{\Sigma}_l^d)^H (\widetilde{\Sigma}_l^d) + N_0 I_M)^{-1} (\widetilde{\Sigma}_l^d)^H\ \hat{y}_l^d \quad (35)$$

This signal for the $l^{th}$ user is then converted to time domain by an $M$-point IDFT operation given by:

$$\hat{x}_l^d = \overline{F}_M\ \hat{\widetilde{x}}_l^d,\ l = 1, 2,..., K \quad (36)$$

where $\overline{F}_M$ is $M$-point inverse IDFT matrix. This signal is then used for decoding of the signal for the $l^{th}$ user.

Now, we will discuss power allocation for all the $K$ users on all the $M$ subcarriers. Considering the $K \times K$ matrix $\alpha_m = diag\ (\alpha_m^1, \alpha_m^2,..., \alpha_m^K)$ in (30), it can be shown that individual information rate (or capacity) for each specific user in downlink can be maximized if [12]:

$$(\alpha_m^i)^2 = (v_m^i - \frac{N_o}{(\Sigma_{m,i}^d)^2})^+, i = 1, 2,..., K \quad (37)$$

where $g^+ = \max(0, g)$. $v_m^i$ is chosen such that:

$$E[\|P_m^i\widetilde{x}_i^d(m)\|^2] \leq E[\|\widetilde{x}_i^d(m)\|^2] \quad (38)$$

## V. SIMULATION RESULTS

The advantage of using SC-FDMA for downlink instead of OFDMA in terms of PAPR and BER performance is analysed in [7]. To validate the inclusion of SIC design for FD system in our architecture, we have carried out MATLAB simulations for BER performance in downlink and uplink. For simulation, we have considered eNB with four antennas ($N_e$=4) and two single antenna UEs ($K$=2). We have used Ghorbani Model [13] for modelling the non-linearity that has been introduced to the complex baseband SC-OFDM symbols. We also add thermal noise (Noise temperature = 290K) to these symbols. For simulations, the channel between each antenna of eNB and UEs' antenna is taken as frequency selective with $L$=7 and uniform power delay profile (UPDP). The modulation scheme used for uplink/downlink is 16-QAM. The bandwidth allocated to the users is taken to be 3MHz. The available bandwidth is split into 256 ($N$) subcarriers, out of which 180 ($N'$) subcarriers are used for carrying data. We used a cyclic prefix of duration 4.69μs. Both the users share all the subcarrier allocated.

For downlink, the performance at UE is considered for received complex SC-FDMA symbols from eNB. In Fig.5, the effect of SIC on the receiver performance at the UE is shown. It is observed that without SIC the receiver has zero throughput. Similar result is obtained when we attempt SIC without taking into consideration the non-linear (NL) distortion components (including the transmit noise). It is observed that when we include the NL distortion components in SIC, the receiver performance is equivalent to half-duplex performance.

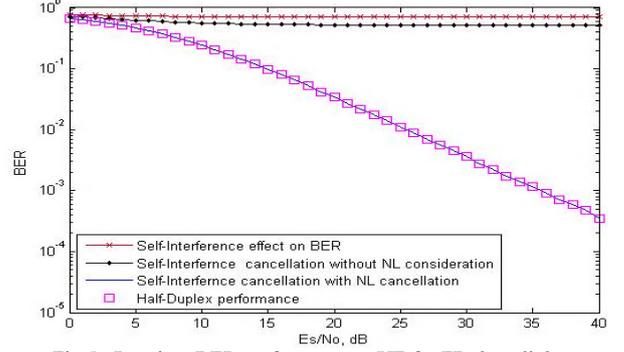

Fig.5: Receiver BER performance at UE for FD downlink

For uplink, the performance at multiple antenna eNB is considered for received complex SC-FDMA symbols from UEs. In Fig.6, the effect of SIC on the receiver performance of the eNB is shown. There is additional diversity gain introduced due to multiple antennas in eNB as discussed in section 4. It is observed that without SIC (which now consists of self and cross talk), the receiver has zero throughput. Similar result is obtained when we attempt SIC without taking into consideration the NL components (including the transmit noise). No improvement in receiver performance is observed by including the NL distortion components in SIC without the cross-talk cancellation (CTC). The receiver performance is equivalent to half-duplex performance when we considered both self and cross talk along with NL distortion components for SIC. Comparing the BER performance at UE and eNB, it can be observed that due to the additional diversity gain, there has been nearly 26 dB gain for the eNB over the optimal BER performance of UE at BER $10^{-3}$.

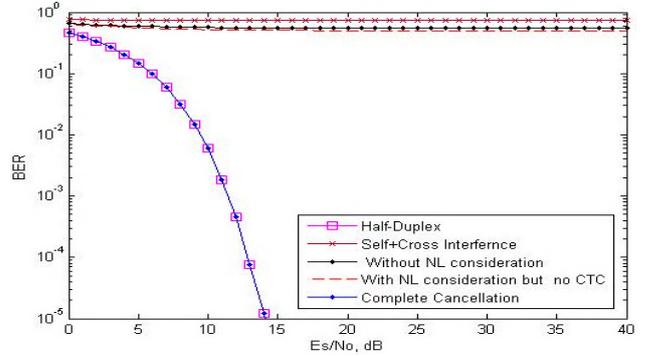

Fig. 6: Receiver BER performance at eNB for FD uplink

## VI. CONCLUSION

In this paper, we propose architecture for full-duplex (FD) multiple antenna eNB and single antenna UE for future 5G networks. The FD operation is expected to result in a decrease in cellular spectrum requirement by half. The SIC design provided in [2,4] for implementation of the FD architecture was briefly discussed. We analyzed the uplink and downlink operation for the proposed FD system. Finally the simulation results validated the inclusion of SIC design for FD uplink and downlink operation.

Acknowledgement

The work is supported by Department of Electronics and Information Technology (DEITY), Government of India.